# Water orientation and hydrogen-bond structure at the fluorite/water interface

Rémi Khatib[1], Ellen H. G. Backus[2], Mischa Bonn[2], María-José Perez-Haro[2], Marie-Pierre Gaigeot[3], and Marialore Sulpizi[1]

[1] Johannes Gutenberg University Mainz, Staudingerweg 7, 55099 Mainz, Germany
[2] Max Planck Institute for Polymer Research, Ackermannweg 10, 55128 Mainz, Germany
[3] LAMBE CNRS UMR8587, Université d'Evry val d'Essonne, Boulevard F. Mitterrand, Bât Maupertuis, 91025 Evry, France

Corresponding author.
E-mail: sulpizi@uni-mainz.de
Phone: +49 6131 3923641

Water in contact with mineral interfaces is important for a variety of different processes. Here, we present a combined theoretical/experimental study which provides a quantitative, molecular-level understanding of the ubiquitous and important $CaF_2$/water interface.
Our results show that, at low pH, the surface is positively charged, causing a substantial degree of water ordering. The surface charge originates primarily from the dissolution of fluoride ions, rather than from adsorption of protons to the surface.
At high pH we observe the presence of Ca-OH species pointing into the water. These OH groups interact remarkably weakly with the surrounding water, and are responsible for the "free OH" signature in the SFG spectrum, which can be explained from local electronic structure effects. The quantification of the surface termination, near-surface ion distribution and water arrangement is enabled by a combination of advanced phase-resolved Vibrational Sum Frequency Generation spectra of $CaF_2$/water interfaces and state-of-the-art ab initio molecular dynamics simulations which include electronic structure effects.

It is now widely accepted that water confined at an interface shows properties on the nanoscale which are remarkably different from bulk water. For water at hydrophobic interfaces, e.g. the water-air interface, the hydrogen-bonded network of water is abruptly interrupted, and non-hydrogen-bonded, "free OH" groups appear, pointing away from the bulk. For water at polar solids, i.e. at mineral surfaces, the charge at the surface as well as the polar groups may be responsible for specific hydrogen bonds and may align the water through strong electrostatic interactions. Water-mineral interactions are of general importance for a wide range of environmental, chemical, metallurgical, and ceramic processes [1, 2]. The interaction of fluorite ($CaF_2$) with water is of specific relevance for industrial, environmental and medical applications, e.g. for understanding fluorine dissolution in drinking water [3].

Recently, the proposal of using $CaF_2$ as an analogue of $UO_2$ in dissolution experiments to understand the long term dissolution behaviour of spent nuclear fuel has renewed the interest in the interaction of this material with water [4].

Despite the apparent importance of the fluorite/water interface, it has been challenging to obtain detailed insights into this interface at the molecular-scale. Nonetheless, Frequency Modulation Atomic Force Microscopy (FM-AFM) [5] has provided important new information on molecular length scales by analysing the fluorite/water interface, not only as function of the pH, but also as function of the concentration of ions in the solution and addressing fluorite/water interfaces with saturated and supersaturated solutions. At high pH, the presence of surface adsorbates is detected, which was attributed to calcium hydroxo complexes [5]. At low pH atomic scale disorder was observed, which could be attributed to either partial dissolution of the topmost layer by the creation of F- vacancies, or to proton adsorption at the interface. Still experiments seem not to be able to distinguish between the two possible scenarios [5]. As another surface sensitive technique, Vibrational Sum Frequency Generation Spectroscopy (VSFG) has the ability to selectively address the nanometric interfacial water layer, and indeed has contributed substantially to our understanding of the physical and chemical properties of the $CaF_2$/water interface [6, 7]. VSFG is rather unique in its ability to provide the vibrational spectrum of water molecules specifically at the interface, as the selection rule of VSFG requires symmetry to be broken, i.e. no VSFG signal can be generated from the adjacent centrosymmetric bulk. Previous VSFG investigations of water at the $CaF_2$/water interface by the Richmond group [6, 7] have revealed dramatic changes in the interfacial hydrogen bonding structure upon changing the pH of the aqueous phase. In particular at low pH, the VSFG experiments have suggested that positive charge develops on the surface, causing orientation of water molecules into highly ordered, tetrahedrally coordinated states. At near-neutral pH, the VSFG signal vanishes and this has been interpreted as the result of a more random orientation of the interfacial water molecules at a near-neutral surface. Finally in the basic pH regime dissociative adsorption was hypothesised to take place on the solid surface resulting in the formation of Ca-OH species. Open questions are still: how do these OH groups contribute to the VSFG spectrum? What type of order is established in the interfacial water region?

Here we present a combined theoretical/experimental study aimed at answering these questions and to provide a new microscopic understanding of the $CaF_2$/water interface as function of pH. We explore the effect of surface termination on interfacial water arrangement and we show the importance of the local electrical field due to ions in solution in the near-surface region on water orientation. Such a detailed analysis is now possible thanks to recent advances in both experimental and computational techniques.

In particular, from the experimental point of view, we move beyond the current state of the art providing the first phase-resolved VSFG spectra for buried $CaF_2$/water interfaces and the first broadband phase-resolved VSFG spectra at the solid-liquid interface in general. In this way we obtain information about the absolute orientation of the interfacial water molecules. Although phase-resolved VSFG [8] and even phase-resolved two-dimensional VSFG, has been successfully applied to water/vapour [9, 10, 11] and water/surfactant interfaces [12, 13, 14, 15], its application to the solid/liquid interface has remained limited to quartz [16] and alumina [17, 18, 19] interfaces. Phase-resolved experimental spectra contain a wealth of spectral information which is crucial to characterise these solid/liquid interfaces and reflect e.g. hydrogen bond strength and water dipole orientation, yet the interpretation of the spectroscopic data in terms of microscopic, atomistic water structures at the interfaces remains challenging, and requires theoretical spectroscopic modelling, which simulations can provide.

From the theoretical point of view, we provide the first atomistic models for such interfaces over a wide range of pHs, thereby allowing for an atomistic interpretation of the experimental spectra. Classical Molecular Dynamics (MD) simulations have already been used to simulate water/air interfaces [20, 21, 22]. However, the parametrization of the different force fields may influence the simulated spectra. To overcome such difficulties, we use Density Functional Theory (DFT) -based molecular dynamics simulations, which allows an accurate description of the structure and dynamics of hydrogen bonding in highly heterogeneous environment, also including electronic polarisation. The VSFG spectra have been simulated thanks to a newly developed approach based on velocity-velocity correlation functions (VVCF). The VVCF have been successfully used in the past to obtain the vibrational modes of solvated systems [23, 24, 25]. In order to reproduce the surface specificity of the VSFG spectroscopy we introduce here specific selection rules which take into account the orientation of the O-H bonds with respect to the incident beams. The newly developed surface sensitive VVCFs permit to considerably accelerate the VSFG signal calculations [26], since they only require the atomic positions/velocities avoiding the additional cost of the calculation of dipole moments and polarizabilities. The obtained speed-up permits to investigate more extensively several models, including e.g. different surface charges.

**Results and discussion**

In the VSFG experiment an infrared and visible laser pulse are in space and time overlapped on the $CaF_2$/water interface and the reflected VSFG signal is detected. In the phase-resolved VSFG experiment a reference signal (local oscillator) is generated from a gold surface before the sample which interferes with the VSFG signal obtained from the $CaF_2$/water interface. By blocking the local oscillator, conventional VSFG spectra can be readily measured, in parallel with the phase-resolved measurements. See the method section for experimental details. The pH of the aqueous solutions is adjusted using concentrated HCl and NaOH solutions for low and high pH, respectively. The solutions are held between two $CaF_2$ windows. The recorded conventional VSFG spectra at pH=2, neutral pH, and pH=13 (without additional salt) are depicted in Fig. 1 top panel, grey curves. At pH=2 an intense signal with a peak maximum around 3200 $cm^{-1}$ is observed.

For neutral and high pH the signals are very weak. At high pH a peak appears at roughly 3645 $cm^{-1}$. These results are in good agreement with literature [6, 7]. The results can qualitatively be understood by noting that at low, resp. high pH, vacancies can be created and a substitution reaction can occur.

When a $CaF_2$ surface is immersed in water, some dissolution is expected. At low pH more of the material will dissolve, because of the excess of hydronium ions. In particular the following reaction is expected to take place, which creates a positively charged fluoride vacancy at the surface:

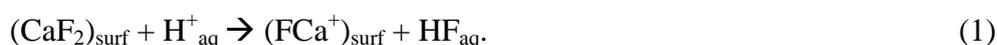

$(CaF_2)_{surf} + H^+_{aq} \rightarrow (FCa^+)_{surf} + HF_{aq}$.          (1)

Hence, at low pH the surface is strongly positively charged aligning the water molecules. As the VSFG signal increases with increasing interfacial order in the system, a large VSFG signal is detected [6, 7].

At high pH, hydroxide ions are present in excess, and are expected to react with the $CaF_2$ surface. In this case, an anion substitution occurs according to:

$$(CaF_2)_{surf} + HO^-_{aq} \rightarrow (FCaOH)_{surf} + F^-_{aq} \qquad (2)$$

As a result, at high pH Ca-OH is possibly formed, which could account for the narrow band signal at 3645 cm$^{-1}$ [6, 7]. At neutral pH the surface carries little charge as only a minor amount of $CaF_2$ will dissolve and the water molecules are disordered and therefore hardly visible in the VSFG spectrum.

These qualitative conclusions, while plausible, have not been verified and do not provide a quantitative explanation for the observed signals. One of the open questions is how many vacancies and how many substitutions there are, at the different pHs. Moreover, as the Ca-OH should be in direct contact with adjacent water molecules, one might not expect such a high frequency for this mode. We will therefore further investigate the hydrogen bonding of such a Ca-OH moiety.

To investigate this assignment in more detail and to obtain information about the orientation of the water molecules, phase-resolved VSFG spectra, which independently provide the real and imaginary parts of the second-order nonlinear response are measured. In particular, the imaginary part of the $\chi^{(2)}$ response constitutes the interfacial equivalent of the bulk absorption spectrum, with the distinction that Im$\chi^{(2)}$ can be both positive or negative, depending on the orientation of the transition dipole moment. In Fig. 1, Re$\chi^{(2)}$ and Im$\chi^{(2)}$ are depicted together with the intensity spectrum inferred from the phase-resolved measurement by combining Re$\chi^{(2)}$ and Im$\chi^{(2)}$. We find perfect agreement between the inferred spectrum and the conventional VSFG intensity measurements, lending credence to the experimental approach. Fig. 1 reveals that Im$\chi^{(2)}$ at pH=2 shows a broad band in the hydrogen bond region which is peaked around 3300 cm$^{-1}$. The negative sign of Im$\chi^{(2)}$ shows that the water molecules which are responsible for this band are collectively oriented with their dipole pointing toward the bulk, in line with the surface bearing positive charge.

The imaginary part of the response for high pH shows a broad negative band which peaks at 3630 cm$^{-1}$. The results confirm the presence of a high-frequency OH group, as was observed in the intensity VSFG spectra.

The phase sensitive experiments show that this high-frequency OH band has a negative sign, showing that the OH oscillators responsible for this band must point from the surface into the bulk. For neutral pH the overall signal is very weak, especially if compared with the low pH regime. (See Fig. 1). In this particular spectrum a positive and negative feature at low and high frequency, respectively, are observed. Note that the exact shape of the spectrum at neutral pH is critically depending on the experimental condition, as we are around the point of zero charge (PZC) [6]. Small fluctuations in the pH can lead to a slightly positive or slightly negative surface charge changing the sign of the imaginary part of the spectrum. Moreover, a different amount of carbon dioxide present in the sample could even change the PZC by several pH units [27]. Nevertheless, the intensity is always very small. The real part of the response at different pH shows, in line with the imaginary part, low signal intensities for neutral and high pH, and a large signal for low pH.

To relate the observed features to a molecular description of the interface, we have modelled the different cases. For low pH, model systems which resemble the final equilibrium state can be built with various concentrations of fluorite vacancies on the surface, which correspond to different extents of positive charge on the surface (Fig. 2). In particular our model consists of a CaF$_2$ slab in contact with water where two equivalent interfaces are present. Fluoride counterions are added to the solution to compensate the positive surface charge, i.e. to get an overall neutral system. We find that the F$^-$ ions tend to prefer to be solvated by water, and form a diffuse layer in the near-surface region. Overall, the surface-localised positive charge and the near-surface negative counterions generate a double layer, giving rise to a rather strong electrical field at the solid/liquid interface.

We simulated an extreme condition with 2.58 vacancies.nm$^{-2}$ (4 vacancies on each surface) and milder conditions with 1.29 vacancies.nm$^{-2}$ (2 vacancies) or with 0.64 vacancies.nm$^{-2}$ (1 vacancy). The spectra for the different vacancy densities are reported in the top row of Fig. 3. The spectral responses are calculated from the surface sensitive vibrational density of states using surface specific VVCF. (see method sections for details). For all the different concentrations of surface vacancies the common feature is the presence of a broad negative band in the Im$\chi^{(2)}$ spectrum, which, for the 1 and 2 vacancies systems, is located around 3300 cm$^{-1}$. As the charge concentration increases to 4 positive charges, the intensity of the band increases and the band position moves towards lower frequencies, with a maximum located at 3100 cm$^{-1}$ which is clearly not consistent with the experiment. Additional information can also be extracted from a comparison between the calculated and experimentally measured Re$\chi^{(2)}$. The computed Re$\chi^{(2)}$ (Fig. 3, blue lines) shows two main peaks, a positive peak at higher frequencies and a negative one at lower frequencies. In the case of 1 or 2 positive charges on the surface the peak position and the crossing from positive to negative values are in good agreement with the experimental spectra (Fig. 1). However as the defect number increases to 4, we notice a very strong shift of the negative band to lower frequencies which also shifts the zero crossing toward 3200 cm$^{-1}$. Moreover, also intensity spectrum matches the best with the experiment for 1 or 2 vacancies per site: the intensity spectrum for 4 vacancies has the maximum at 2800 cm$^{-1}$ which is far too low. This suggests that the vacancy density is around 0.65 per nm$^2$ for the experimental condition of pH=2.

A detailed molecular analysis unveils the molecular structure which is responsible for the strong negative band in Im$\chi^{(2)}$. Indeed such a band is due to an ordered layer of water which builds up at the interface, with water dipoles oriented toward the bulk. As is clear from Fig 4a, showing the Im$\chi^{(2)}$ spectrum with increasing probing thickness, the order extends over 4-5 Å. Including water molecules further than 5 Å from the surface does not change the shape or the intensity of the calculated VSFG spectrum. Remarkably, even for a strongly charged interface the aqueous order only extends over 4-5 Å, which corresponds to roughly 2-3 layers of water. We should note here that the high computational cost of electronic structure based methods imposes severe limitations on the size of the accessible models. In this respect our model is expected to capture the contribution to the spectra of the Stern layer (possibly the major contribution here), but cannot account for the full diffuse layer, which is expected to extend over a few nanometers thickness. The experimentally estimated Debye length for CaF$_2$ is roughly 30 Å [28].

As mentioned in the introduction, one of the suggested interpretation for the atomic scale disorder observed at low pH in the FM-AFM experiments, is proton adsorption at the interface [5]. In order to investigate such a scenario, we build an additional model where no fluorine vacancies are present, but instead excess proton in the form of dissociated HCl is present (4 HCl, solution 2.5 M). Such a system would eventually corresponds to 2 excess proton per surface. The proton distribution at the interface is reported in Supplementary Fig. 1. The idea here is to isolate the contribution to the VSFG spectra coming from the excess proton only. The calculated VSFG spectra for this system are shown in Fig. 3 as dotted lines. Overall the signal is much weaker than those characterising the model with two fluorine vacancies per surface. Moreover, the peak in the Im$\chi^{(2)}$ is located at roughly 3500 - 3600 cm$^{-1}$, which is certainly far from the experimental peak location. From this analysis we can conclude that the excess proton alone cannot be responsible for the measured spectra, which instead originates from the water aligned by the positive fluorine vacancies.

For high pH, we have constructed a model where a surface modification of the $CaF_2$ has taken place in response to the increased concentration of OH groups in the solution. In the topmost fluorite layer, $F^-$ were partially or totally replaced by $HO^-$ (Fig. 2). Different concentrations of OH have been considered in order to establish a relation between the VSFG signal intensity and the pH: 1, 6 and 12 substitution over the 12 available sites per surface.

The imaginary and real part of the VSFG spectrum together with the intensity spectrum calculated from the surface selective VVCF analysis are presented in the bottom row of Fig. 3 for the three different values of OH concentration on the surface. For the 1 and 6 substitutions two main features can be observed in the imaginary part: the first is a positive band between 3280 and 3400 $cm^{-1}$, the second is a negative feature between 3400 and 3700 $cm^{-1}$. In the case of the maximum number of 12 OH defects, the overall profile of $Im\chi^{(2)}$ is very different, with a broad negative band extending up to 3200 $cm^{-1}$ where a crossing to positive values is finally observed. The real part and the intensity spectrum exhibit for the case with 12 defects a very high intensity below 3600 $cm^{-1}$, which is not present in the experiment (Fig. 1). If we compare the calculated spectra with the experimental ones we can observe that the 1 or 6 OH substitution are in quite good agreement with the experiments. Therefore at pH=13 we can set an upper limit of 6 OH substitutions corresponding to 3.87 substitutions.$nm^{-2}$ per site.

When decomposing the overall signal in terms of molecular contributions we can provide a microscopic interpretation of the spectra. In particular we can show that the peak between 3600 and 3700 $cm^{-1}$ is only associated with the OH groups on the surface, namely those OH groups which replace $F^-$ in the topmost layer, which is clear from the purple spectrum in the bottom panel of Fig. 4. This frequency is very close to that of "free OH" [29, 30] and it can indeed be verified that such an OH group on the surface does not hydrogen bond to water. This is clearly shown in the radial distribution function of the Ca-OH hydrogen with water oxygens; the distance between the proton of the Ca-OH and the oxygen from water (red curve, Fig. 5) is much larger than the distance between the proton from one water molecule and the oxygen from the next water molecule (black curve Fig. 5). A similar peak at high frequency has also been observed for the alumina surface [31], where such hydrogen bond is not formed between the surface OH group and the water molecules.

The peak between 3280 and 3400 $cm^{-1}$ is instead associated with hydrogen bonded water molecules at the interface. Their orientation is opposite to that of the OH groups (as evident from the opposite sign of $Im\chi^{(2)}$ for the two different peaks). This ordering is not very pronounced and saturates with a distance of 2 Å. This positive peak is hardly observed in the experiments, possibly caused by a different pH compared to the modelling or due to a phase uncertainty (see experimental methods). Please note that it has been shown that in experiments the $CaF_2$ interface can become negative at high pH due to the conversion of carbon dioxide into carbonate [32]. Subsequently, carbonate can bind to the surface. To exclude that the 3630 $cm^{-1}$ peak in the experimental spectrum originates from OH groups from species like bicarbonate, we measured VSFG spectra for a 0.1 M NaOH and 5 mM $Na_2CO_3$ solution. We observe that the 3630 $cm^{-1}$ peak decreases upon adding carbonate to the solution showing that the peak does not originate from the carbonate. The addition of carbonate could shift the substitution reaction of $F^-$ for $HO^-$ (equation (2)) reducing the $HO^-$ amount on the surface. Moreover, also upon adding $CaF_2$ to the solution at low and high pH we observe a reduction of the VSFG signal, which could be explained by less $CaF_2$ dissolution (equation (1)) at low pH and a different substitution equilibrium at high pH. Screening of the surface charge by adding salts could result in a lower signal at low pH.

For neutral pH we use in the model a fluorine terminated surface in contact with neutral water (no excess of hydronium or hydroxide). The calculated $Im\chi^{(2)}$ is reported in Fig. 1 (blue line), the overall signal intensity (Fig. 1, black line) is very weak and presents a negative sign in the higher frequency region (3400-3500 cm$^{-1}$) and a positive band in the lower frequency range (3000-3200 cm$^{-1}$). A molecular analysis shows that, surprisingly, there is a strongly adsorbed layer of water with, however, little to no preferential orientation at the interface.

By comparing the calculated and experimental signal intensities for different pH, a more precise estimation of the different vacancies or substitutions at different pH can be made. For high pH we concluded above that the 1 or 6 substitutions match the experiment very well. For low pH we concluded that 1 or 2 vacancies are in agreement with the experimental spectra. Based on the relative intensity between low and high pH, we can refine our conclusion. As can be seen in Fig. 1 the experimental and calculated spectra are very well matched with 1 vacancy per surface (0.65 vacancies.nm$^{-2}$) for at pH=2 and with 6 substitutions per surface (3.87 substitutions.nm$^{-2}$) at pH=13.

**Conclusions**

We present a combination of phase sensitive VSFG and *ab initio* molecular dynamics modelling which permitted to elucidate the details of the fluorite/water interface. The calculated VSFG spectra using surface selective VVCFs provide a molecular assignment of the different features observed in the experimental spectra. We find that at low pH the strong band in the hydrogen bond region is due to the highly ordered water as the surfaces is positively charged, due to the F$^-$ dissolution. We also show that an eventual excess proton at the interface can only have a minor impact on the spectra. At high pH the "free OH" signal is due to the surface Ca-OH groups, which do not hydrogen bond strongly to water. The very good agreement between theory and experiments in both the Re$\chi^{(2)}$ and Im$\chi^{(2)}$ permits to pin down the atomistic details of the CaF$_2$ interface with water and to provide a first molecular interpretation of the spectra.

**Methodology**

<u>1 VSFG setup</u>

For the VSFG experiments a 5 mJ Ti:Sa amplified laser (Spitfire Ace, centred at 800 nm with a pulse duration of roughly 40 fs) has been used. Roughly 2 mJ of the laser output is used to pump an optical parametric generation/amplification stage (TOPAS, light conversion) to generate pulses around 3300 cm$^{-1}$ and a bandwidth of 500 cm$^{-1}$. 1 mJ of the laser output is passed through an etalon resulting in the narrow band visible (VIS) pulse giving the roughly 25 cm$^{-1}$ spectral resolution of the experiment. At the sample the IR and VIS pulse have both an energy of roughly 3 µJ. After focusing with a 100 cm and 5 cm focal length lens for the VIS and IR, respectively, both beams are overlapped in space and time on a gold mirror at relative grazing incidence to generate the local oscillator. All three beams are refocused by a curved mirror with a focal length of 50 mm and guided to the sample where the IR and VIS have an incidence angle of roughly 45 and 40 _ with respect to the surface normal, respectively. The local oscillator is delayed with a 1 mm thick fused silica plate. Subsequently, the local oscillator and the VSFG signal from the sample are dispersed in a spectrometer and detected with an EMCCD camera.

<u>2 Samples</u>

The low pH solution has been prepared by diluting 12 M HCl with millipore water to 0.01 M resulting in a pH of roughly 2 which has been checked with pH paper. A solution with a pH of roughly 13 has been obtained by diluting NaOH in millipore water at a concentration of 0.1 M. A few drops of the aqueous solution are held in between two CaF$_2$ windows with a random surface termination. Great care is taken that the CaF$_2$ is reproducibly placed in the same way in the VSFG setup; we have noticed variation in primarily the signal intensity if the CaF$_2$ plate is rotated around its surface normal indicating that there is some anisotropy of the sample. The CaF$_2$ windows are cleaned on a daily basis by annealing them 2 hours in an oven at 500 °C. In between different experiments on one day the windows are thoroughly rinsed with Millipore water. Following this cleaning procedure, no C–H contamination is observed in the VSFG spectra.

3 Data analysis

To experimentally obtain the real and imaginary parts of the VSFG response and to correct for the frequency dependence of the IR light, we measured a conventional and a phase sensitive VSFG spectrum for the CaF$_2$/gold interface under the same conditions as for the CaF$_2$-aqueous interface. Therefore we coated the CaF$_2$ window used to obtain VSFG spectra from the aqueous interface with a 100 nm thick gold layer without a chromium layer (this is often used to enhance the adhesion of gold to substrates). The VSFG spectra are measured such that, as in the real experiment the IR and VIS beam, and subsequently the VSFG beam as well, pass through the CaF$_2$ window. Care is taken that the sample lies flat and that the height is identical for every aqueous sample and the gold sample by obtaining the VSFG signal always at the same height on the CCD camera. The reflection from the CaF$_2$/water interface for 632 nm is unfortunately too weak to enable height measurements with a height sensor, as has previously been done for phase-resolved measurements from the water-air interface [13, 11]. Covering a part of the CaF$_2$ surface with gold to enable reflection of the 632 nm is impossible as the gold slightly dissolves resulting in the deposition of gold particles on the bare window. The procedure to obtain the imaginary and real part by Fourier transformation, selecting the appropriate term, and deviation by the gold reference spectrum has been explained in detail in [13, 11]. Subsequently the phase has to be correct by 170 ° due to differences in the reflectivity and Fresnel factors for the CaF$_2$/gold interface and the CaF$_2$/water interface, because the refractive index of gold is complex. A thin (1-2 nm) layer of chromium between the CaF$_2$ and gold film changes this phase correction dramatically and should therefore not be used. The phase uncertainty due to fluctuations in the sample position is estimated to be around 30 °. Rephasing with a spectrum from the CaF$_2$/D$_2$O interface, as is usual used for the water-air interface, [13, 11] is not possible here, as the nonresonant VSFG signal from this interface is too weak.

4 Fresnel factors

As is common in the VSFG community the measured VSFG spectra are not corrected for the Fresnel factors. We decided to multiply for the comparison in Fig. 1 the calculated imaginary and real part (and the intensity spectrum as well) of the VSFG response with the Fresnel factors to allow for direct comparison with the experiment. Please note that the calculated spectra in the other figures are purely the nonlinear susceptibility $\chi^{(2)}$. The Fresnel factors are calculated according to [33] using the refractive index of water for the interfacial refractive index [34]. Moreover, the refractive index of CaF$_2$ and water are obtained from [35] and [36], respectively.

5 Simulation Setup

Several models are used to describe the fluorite/water interface over a wide range of pH. The reference system – an interface between CaF$_2$ (111) and water at neutral pH – is composed of 88 water molecules and 60 formula units of CaF2 contained in a 11.59 Å × 13.38 Å × 34.0 Å cell periodically repeated in the (*x*, *y*, *z* ) directions. All the other models have close compositions and size to allow inter-system comparisons. The thickness of water slabs is around 20 Å along the *z* -axis, which is reasonable compromise between the need to achieve bulk-like properties far from the surface and the computational cost. Simulations were carried out with the package CP2K / Quickstep [37], consisting in Born-Oppenheimer MD (BOMD) BLYP [38, 39], electronic representation including Grimme (D3) correction for dispersion [40], GTH pseudopotentials [41, 42], a combined Plane-Wave (280 Ry density cutoff) and TZV2P basis sets. All the BOMD are performed using the NVT ensemble. The Nosé-Hoover thermostat is used to control the average temperature at 330K. Trajectories are accumulated for at least 50 ps (whom 10 ps of equilibration) with a time step of 0.5 fs.

6 Method for VSFG The starting equation to calculate the VSFG response function from molecular dynamics simulations have been introduced by Morita [43, 44, 45, 46]:

$$\chi_{PQR}^{(2),R} = \frac{-i}{k_B T \omega} \int_0^{+\infty} e^{i\omega t} \left\langle \dot{A}_{PQ}(t) \dot{M}_R(0) \right\rangle dt$$

(3)

Here $\chi^{(2),R}$ is the resonant part of second-order susceptibility tensor, (P,Q,R) are any directions of the laboratory frame ($\hat{X}$, $\hat{Y}$, $\hat{Z}$), $\omega$ is the frequency of the IR beam, $A_{PQ}$ and $M_R$ are respectively the components of the total polarizability tensor and the total dipole moment and the dot stands for the time derivative.

If we suppose that at the frequencies of interest only the O-H stretching has an impact on the spectra, the total polarizability and dipole moment of the system ($A_{PQ}$, $M_R$) can be decomposed into individual (OH) bond contributions ($\alpha_{mn,PQ}$, $\mu_{mn,R}$), where the sum is done over all the $N_m$ bonds of the M molecules:

$$\begin{cases} \dot{A}_{PQ}(t) = \sum_{m=1}^{M} \sum_{n=1}^{N_m} \dot{\alpha}_{mn,PQ}(t) \\ \dot{M}_R(t) = \sum_{m=1}^{M} \sum_{n=1}^{N_m} \dot{\mu}_{mn,R}(t) \end{cases}$$

(4)

Moreover, thanks to basic geometry considerations, one can express the dipole moment of the A-B bond from the molecular frame ($\mu_b$) to the laboratory frame ($\mu_l$):

$$\mu_l = D\mu_b$$ 
(5)

where D is the direction cosine matrix projecting the molecular frame onto the laboratory frame. In the following, we will assume that (1) the bond elongations are small enough to make Taylor expansion at the first order and (2) the stretching mode of the bond is much faster than the modes involving a bond reorientation − for example the libration. The second assumption means that $D_{Ri} \sim 0$ and that $dr_z/dt \gg dr_x/dt \sim dr_y/dt$. Therefore $\mu_R$ can be simplified into:

$$\begin{aligned}
\dot{\mu}_R(0) &\approx \sum_i^{x,y,z} D_{Ri}(0) \dot{\mu}_i(0) \\
&\approx \sum_i^{x,y,z} D_{Ri}(0) \left( \sum_j^{x,y,z} \frac{d\mu_i}{dr_j} \frac{dr_j}{dt} \bigg|_{t=0} \right) \\
&\approx \sum_i^{x,y,z} D_{Ri}(0) \frac{d\mu_i}{dr_z} v_z(0)
\end{aligned}$$

(6,7,8)

where $v_z = \frac{dr_z}{dt}\big|_{t'=0}$ corresponds to the projection of the velocity on the bond axis.

With the same methodology for the polarizability, one deduces that:

$$\dot{\alpha}_{PQ}(t) \approx \sum_i^{x,y,z} \left[ D_{Pi}(t) \sum_j^{x,y,z} \left( \frac{d\alpha_{ij}}{dr_z} D_{Qj}(t) \right) \right] \times v_z(t) \quad (9)$$

The use of equation (8) and (9) into equation (4) brings important computational advantages. Indeed the velocities and the direction cosine matrix ($v_z$, D) can be readily obtained from the DFT-MD trajectories while $d\alpha_{ij}/dr_z$, $d\mu_i/dr_z$ can be parametrized [47]. Our approach avoid the additional direct calculation of the bond dipole moment and polarizabilities which, at an *ab initio* level certainly requires a considerable additional computational cost, e.g. the cost of the Wannier centres localisation [48]. Finally, with the splitting of the dipole moment and polarizability into their bond contributions, it is easy to decompose the signal into its auto-, intramolecular and intermolecular parts.

The parametrization of $d\alpha_{ij}/dr_z$ and $d\mu_i/dr_z$ is based on the calculation of the maximally localised Wannier functions (MLWF) [49] and has been done through the methodology developed by Salanne *et al.* [50]. The values are obtained by a 2-point numerical differentiation: a single O-H bond is elongated by • ±0.02 Å. For the O-H bond of water molecules, a trajectory of 128 H2O inside a cubic box (c = 15.6404 Å) has been simulated and an average involving more than 4000 bonds distributed over a dyna mic of 40 ps has been done. One formula unit of HCl has been added to the previous box in order to do the same kind of sampling about the O-H bond of the hydronium. Finally, for the O-H bond of the grafted hydroxide ions, the derivatives are those obtained on a linear monomer of FCaOH. All these values are resumed in the Table 1.

**Acknowledgement**

This work was partially supported by an ERC Starting Grant (Grant No. 336679) and by the European Union Marie Curie Program (CIG Grant No. 334368). All the dynamics were simulated on the supercomputers of the High Performance Computing Center (HLRS) of Stuttgart. The authors would like to thank Y. Nagata and J. Hunger for useful discussion.

**Author contributions**

R.K. did the simulations, the figures and drafted the manuscript. E.H.G.B. designed the study, performed the experiments and wrote the manuscript. M.B. wrote the manuscript. M.-J.P.-H. performed the experiments. M.-P.G. proposed the idea of VVCF. M.S. designed the study and wrote the manuscript. All authors read and approved the final manuscript.

**Competing financial interests**

The authors declare no competing financial interests.

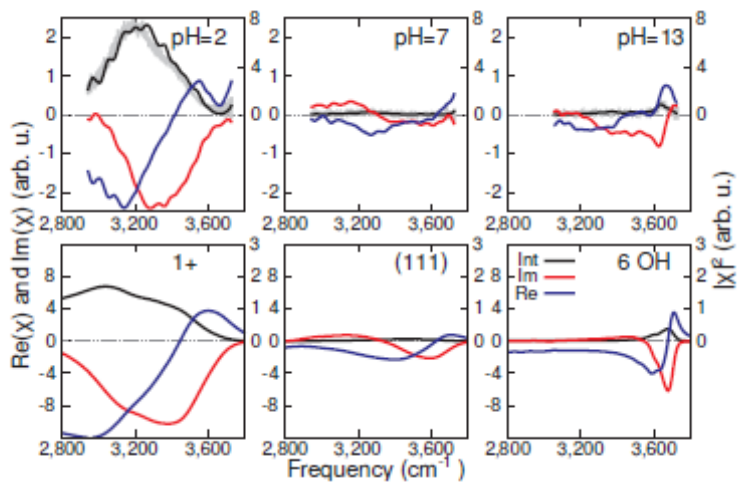

Figure 1: Comparison between experiments and theory. Top: experimental conventional VSFG intensity spectra (grey) and intensity spectra obtained from phase-resolved experiments (black). $\mathrm{Im}\chi^{(2)}$, $\mathrm{Re}\chi^{(2)}$ are also shown as blue and red lines respectively. In the lower panel the analogous quantities ($\mathrm{Im}\chi^{(2)}$, $\mathrm{Re}\chi^{(2)}$, $|\chi^{(2)}|^2$) are reported with the same colour coding. pH 2, 7 and 13 are considered (left, middle, right). In the calculated $\mathrm{Im}\chi^{(2)}$, $\mathrm{Re}\chi^{(2)}$, $|\chi^{(2)}|^2$ the Fresnel factors have been taken into account in order to allow a straightforward comparison to the corresponding experimental spectra.

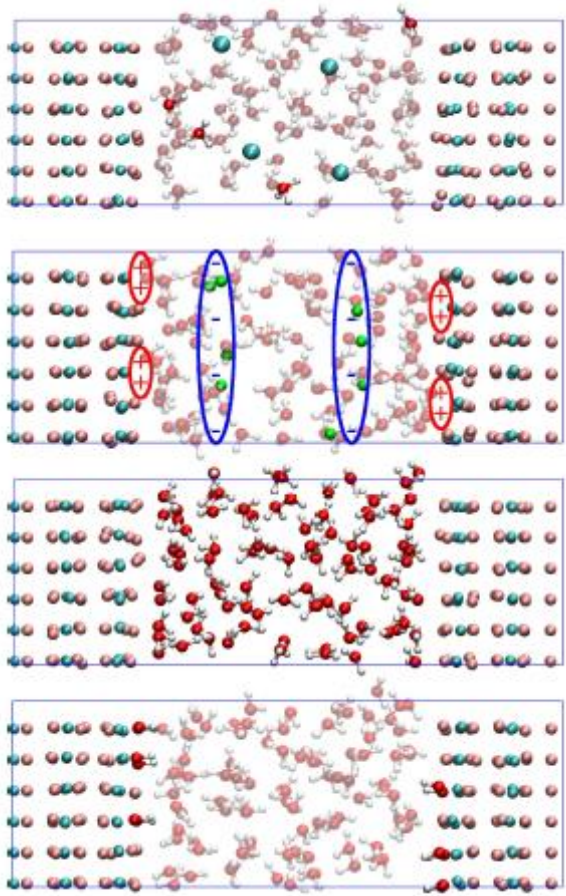

Figure 2: Random snapshots of the system used to describe the CaF$_2$/H2O interface for low (one cell containing 4 HCl, the other one representing the dissolution of the topmost fluorite layer), neutral and high pH. Water molecules are transparent in order to highlight the vacancies, substitutions and the fluoride ions in solution (green).

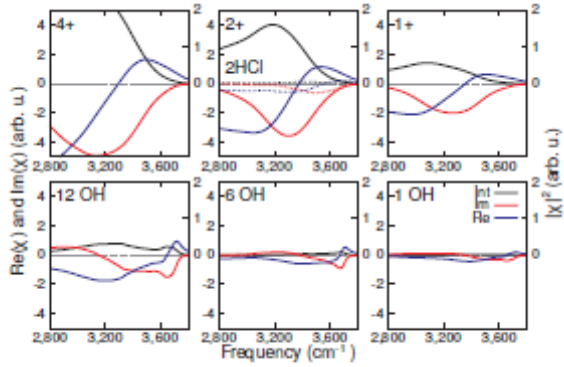

Figure 3: Comparison of the Im$\chi^{(2)}$, Re$\chi^{(2)}$ and $|\chi^{(2)}|^2$ for different values of the surface defect concentration (plain lines). Top panels: low pH. Bottom panels: high pH. In order to facilitate the comparison, the spectra with 2 HCl per surface have been plotted in dotted lines on the spectra with 2 vacancies per surface.

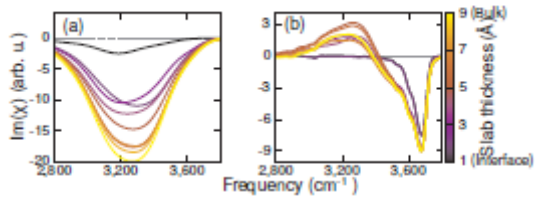

Figure 4: Im$\chi^{(2)}$ as function of the layer thickness included in the calculation. Left panel: low pH (1 defect per surface); Right panel: high pH (6 substitutions per surface).

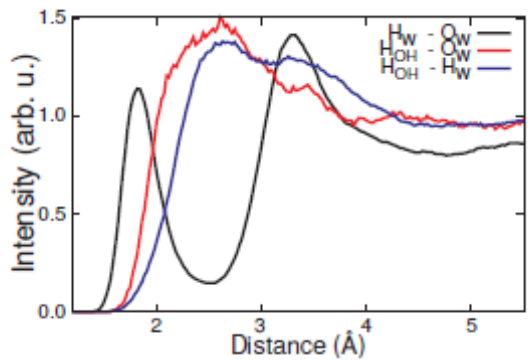

Figure 5: HO and HH radial distribution functions. The subscript "W" stands for water, while the subscript "OH" the grafted hydroxide.

Table 1: Dipole moment (D.Å−1) and polarizability (Å2) derivatives

| | $\frac{\partial \mu_x}{\partial r}$ | $\frac{\partial \mu_y}{\partial r}$ | $\frac{\partial \mu_z}{\partial r}$ | $\frac{\partial \alpha_{xx}}{\partial r}$ | $\frac{\partial \alpha_{yy}}{\partial r}$ | $\frac{\partial \alpha_{zz}}{\partial r}$ | $\frac{\partial \alpha_{xy}}{\partial r}$ | $\frac{\partial \alpha_{xz}}{\partial r}$ | $\frac{\partial \alpha_{yz}}{\partial r}$ |
|---|---|---|---|---|---|---|---|---|---|
| $H_2O$ | -0.15 | 0.0 | 2.1 | 0.40 | 0.53 | 1.56 | 0.0 | 0.02 | 0.0 |
| $H_3O^+$ | -0.11 | 0.0 | 1.7 | 0.47 | 0.40 | 1.50 | 0.0 | 0.0 | 0.0 |
| $HO^-$ | 0.0 | 0.0 | 1.6 | 0.5 | 0.5 | 2.3 | 0.0 | 0.0 | 0.0 |

Calculated derivatives of the dipole moment (D.Å$^{-1}$) and polarizability (Å$^2$) of the O-H bond in a bulk of water and in FCaOH monomer. The results are given within the bond frame.

Supplementary Figure 1: Proton distribution across the interface

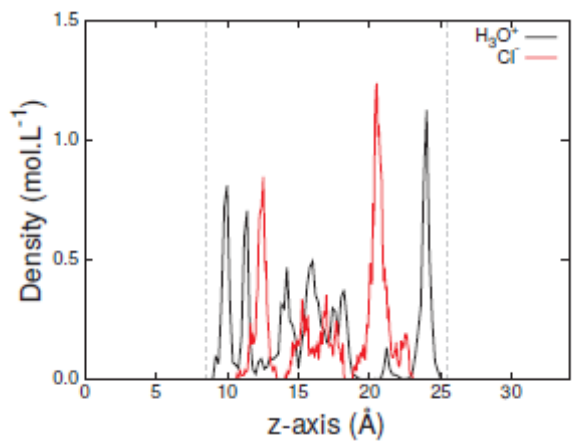

Density profile of $H_3O^+$ and $Cl^-$ along the z-axis. As a guide for the eyes, the position of the $CaF_2$ interface is represented by a dashed grey line.

The density of hydronium ($H_3O^+$) and $Cl^-$ have been plotted on Supplementary Fig. 1. The $H_3O^+$ has been defined as an oxygen atom surrounded by 3 hydrogens at a distance lower than 1.3 Å. One can notice that the regions with a high density of $Cl^-$ are associated with a low density of $H3O^+$ ions. This is in agreement with the low pKa of HCl and shows that HCl is fully dissociated in water.